\documentclass[12pt]{article}

\usepackage{amsfonts}
\usepackage{amssymb}
\usepackage{amsmath}
\usepackage{verbatim}
\newcommand\be{\begin{equation}}
\newcommand\ee{\end{equation}}
\newcommand\bea{\begin{eqnarray}}
\newcommand\eea{\end{eqnarray}}

\newcommand{\R}{{\mathbb R}}

\newcommand{\Td}{{{\mathbb T}^d}}

\newcommand{\Ed}{{\cal E}_d}
\newcommand{\Edh}{\hat{\cal E}_d}

\newcommand{\lv}{l_\text{v}}
\newcommand{\Et}{E_\text{t}}
\newcommand{\lt}{l_\text{t}}

 \newcommand{\inv}{^{-1}}
 \newcommand{\tran}{^\text{t}}     % choose or up 

\newcommand{\TR}{^{\rm T}}

\begin{document}

\begin{titlepage}

\centerline{\bf \Large AN INFORMAL SUMMARY OF} \vspace*{0.5em}
\centerline{\bf \Large A NEW FORMALISM FOR CLASSIFYING} \vspace*{0.5em}
\centerline{\bf \Large SPIN-ORBIT SYSTEMS USING TOOLS } \vspace*{0.5em}
\centerline{\bf \Large DISTILLED FROM THE THEORY OF BUNDLES\footnote{\rm
 Based on a presentation at Spin2014, The 21st International Symposium on Spin Physics, 
 Beijing, China, October 2014. To be published in the International Journal 
 of Modern Physics, Conference Series.}
 }
\vspace*{1.5em}

\centerline{\large K.~HEINEMANN}
\centerline{\textit{Department of Mathematics and Statistics, University of New Mexico, Albuquerque, New Mexico, USA}} 
\centerline{ {\tt heineman@math.unm.edu}}
\vspace*{1em}

\centerline{\large D.~P.~BARBER}
\centerline{\textit{Deutsches~Elektronen--Synchrotron DESY, Hamburg, Germany}}
\centerline{ {\tt mpybar@mail.desy.de}}
\vspace*{1em}

\centerline{\large J.~A. ~ELLISON}
\centerline{\textit{Department of Mathematics and Statistics, University of New Mexico, Albuquerque, New Mexico, USA}}
\centerline{ {\tt ellison@math.unm.edu}}
\vspace*{1em}

\centerline{\large M.~VOGT}
\centerline{\textit{Deutsches~Elektronen--Synchrotron DESY, Hamburg, Germany}}
\centerline{ {\tt vogtm@mail.desy.de}}
\vspace*{1em}

 \centerline{(January 30, 2015)}

\begin{abstract}
  We give an informal summary of ongoing work which uses tools
  distilled from the theory of fibre bundles to classify and connect
  invariant fields associated with spin motion in storage rings.
  We mention four major theorems. One ties invariant fields with the
  notion of normal form, the second allows comparison of different
  invariant fields and the two others tie the existence of invariant
  fields to the existence of certain invariant sets.
  We explain how the theorems apply to the  spin dynamics of
  spin-$1/2$ and spin-$1$ particles.
  Our approach elegantly unifies the spin-vector dynamics from the
  T-BMT equation with the spin-tensor dynamics and other dynamics and suggests an avenue
  for addressing the question of the existence of the invariant spin field.
 \end{abstract}

\end{titlepage}

\section{\label{sec:intro}Introduction}
The polarization of a beam of spin-$1/2$ or  spin-$1$ particles circulating in a storage ring is best
systematized in terms of invariant spin fields (ISF's) and 
invariant polarization-tensor fields (ITF's). They are essential for
describing the equilibrium polarization state of proton, deuteron and electron beams. 

We have already treated the concept of the ISF in depth in Ref.~\cite{BEH}.
Numerical evidence indicates
that ISF's can be rather complex entities.
Moreover the question of existence, although trivial
in some simple cases, e.g., on orbital resonance, is up to this day, unsolved and, as evidenced by
use of stroboscopic averaging \cite{HH}  model situations can occur off orbital resonance where ISF's 
might not exist. In other
words there is no good understanding of the conditions under
which the ISF exists off orbital resonance.
{Nevertheless we believe that practically relevant
spin-orbit systems which have no    such   ISF are ``rare''.}
We call this the ``ISF-conjecture''. \cite{IPAC14,arXiv1,arXiv2}
This has motivated us to extend our studies using a new approach, namely with 
tools from  Dynamical-Systems theory 
developed in the 1980s by R. Zimmer, R. Feres 
and others \cite{Zi,Fe,HK} and following previous work by one of us (KH) in his PhD thesis.\cite{HE1} 
In particular, this approach enables us to generate, classify and study new and old invariant fields
and it provides an avenue for addressing the question of the existence of ISF's.
{ 
Since we work in the framework of topological dynamical systems, all functions
of interest are continuous, in particular the ISF.}

We have proved  { several major theorems, among which are
the following four.} They are
 the Normal Form Theorem, 
tying invariant fields with the notion of normal form,
 the Decomposition Theorem, allowing comparison of different invariant fields,
 the Invariant Reduction Theorem, giving new insights into 
the question of the existence of invariant fields
%(and in particular invariant spin fields) 
and which is supplemented by the Cross Section Theorem.
It turns out that the { well-established notions 
\cite{BEH}} of
invariant frame field {and uniform invariant frame field}
are generalized by the normal form concept whereas
the well-established notions of ISF and ITF
are generalized to the invariant
$(E,l)$-fields to be defined later. 
With the flexibility in the choice of $(E,l)$,
we have a unified way to study, for example, the dynamics of spin-$1/2$ 
and spin-$1$ particles and the density matrices of the 
{ corresponding} particle bunches. 

Our formalism, which we call the Technique of Association (ToA), has
its origins in Bundle Theory so that our methods can be compared with
those used in Yang-Mills Theory.  We thus open a significant new area
of research in our field by bringing in techniques from Bundle Theory
used hitherto in very different research areas.  We believe that all
four of these theorems are important. Furthermore, we come to a new view of the ISF via its link to the existence of 
a certain invariant set.

Owing to the page limit, this account in necessarily just an {\em informal summary} of this work with which we wish 
to increase its 
accessibility for accelerator physicists. We hope that the reader will be motivated to 
look at the complete and  rigorous accounts of the mathematics in Refs.~\cite{IPAC14}--\cite{arXiv2}
where important notation and terminology are also explained.
%\section{\label{sec:landau}The Theoretical Minimum (c.f. Lev Landau)}
%Moreover we must assume that 
%%We assume that 
%the reader is familiar with concepts like
%the T-BMT equation for spin motion, the 
%ISF, the invariant frame fields (IFF), and the amplitude dependent spin tune (ADST).
\vspace{-0.0mm}
\section{\label{sec:basic}The Basic Equations}
We assume that the reader is familiar with the T-BMT equation for spin motion, the invariant spin field (ISF) and
the amplitude-dependent spin tune (ADST).\cite{BEH}

In contrast to Ref.~\cite{BEH}, we now work in terms of 1-turn maps for the orbital and spin motion.
The 1-turn spin map from the T-BMT equation starting at azimuth $\theta_0$ with orbital phases $\phi(\theta_0)$ and with
amplitudes $J$ is
$${  S(\theta_0 + 2 \pi)  = A(\theta_0, \phi(\theta_0), J ) S(\theta_0)}$$
where $S$ is the column matrix of the three components of the 
 { spin vector} and where  the 1-turn spin map {  $A(\theta, \phi, J)$} 
is  { a  real orthogonal  {  $3 \times 3$} matrix of unit 
determinant, i.e.,}
an element of  {  $SO(3)$}, and it is a {\em {function}} of { $(\theta, \phi, J)$},
{ $2 \pi$-periodic in   $\theta$ and the components of $\phi$}.
Over one turn, the position $\phi$
on the  { $d$}--torus { $\Td$} is transformed to
 {$j_{J}(\phi)$}. 
%%%%%%%%%%%%{\color{green} where $j_{J}(\phi)$ is $2\pi$-peropdic in $\phi$.} 
Usually $j_J(\phi) = \phi + 2\pi \omega(J)$
% with orbital tunes $\omega$.
where $\omega(J)$ is the set of orbital tunes. 
%but this can be generalized.
%.However, the use of $j_J$ allows for generalisations.   
Normally { $d = 1, 2$} or { $3$}.
We ignore the ambiguous and miniscule Stern-Gerlach forces.
For our integrable orbital motion $J$ is a constant parameter which we now suppress.
\vspace{-0.0mm}
\section{\label{sec:SF}The Invariant Spin and Tensor Fields}
As explained in Refs.~\cite{BEH} and \cite{arXiv2} the maximum attainable vector polarization of a beam of  particles of arbitrary spin is
expressed in terms of the invariant spin {\em field} (ISF). 
%${f_v}$.  
%{ ($\equiv  \hat n(\theta, \phi,J)$)} 
{ An ISF, $f_v(\theta,\phi)$, being a spin field, evolves as a consequence of the T-BMT equation so that 
after one turn $f_v(\theta + 2 \pi, j(\phi)) = A(\theta, \phi) f_v(\theta, \phi)$. By definition $|f_v| = 1$ and $f_v$ is $2\pi$-periodic 
in $\theta$, i.e.,  $f_v(\theta + 2 \pi,\phi) = f_v(\theta, \phi)$. Then $f_v(\theta, j(\phi) ) = A(\theta, \phi) f_v(\theta, \phi)$.}
Note that in Refs.~\cite{BEH} and \cite{BV2} and in most other literature on ISF's, the ISF is written as  $\hat n (\theta,\phi)$.

In Refs.~\cite{IPAC14}--\cite{arXiv2} and this work we ``sample'' the particle and spin motion once per turn at some fixed 
azimuth $\theta_0$. So we suppress the $\theta_0$ and write $f_v(j(\phi)) = A(\phi) f_v(\phi)$.

For deuterons (spin-1) a full description of the equilibrium polarization state of a beam involves the 
invariant polarization-tensor field (ITF), $f_t$. This is a {$3\times3$}, real, symmetric, traceless,  
field on $\Td$ which we also take to be continuous in $\phi$.
 { The invariance of
$f_t$ over one turn gives}: $f_t(j(\phi)) = A(\phi) f_t(\phi) A(\phi)^T$.
%The ITF  { {$T^I$}} ({\magenta{$f_T$}}):  { {$$T^I(j(z)) = A(z) T^I(z) A(z)^T$$}}. As shown in 
As shown in Ref.~\cite{BV2}  { one may parametrize $f_t$ as} 
$f_t  = \pm   {\sqrt{\frac{3}{2}}} \left \{ \hat n {\hat n}\TR  - \frac{1}{3} I \right \} 
(\equiv \pm   {\sqrt{\frac{3}{2}}} \left \{ f_v {f^T_v}  - \frac{1}{3} I \right \} ) $.
%where  we use the symbol $\hat n$  (instead of $f_v$) used there.
\vspace{-0.0mm}
\section{\label{sec:unify}A Unified Representation of the Transformations}
We can encompass the  { varieties of} ``spin'' 
dynamics in  $SO(3)$-spaces, $(E,l)$,  where $E$ is a
topological space and $l$ is a continuous  $SO(3)$--action on $E$,
i.e., $l:SO(3)\times E\rightarrow E$ is continuous and
$l(I;x)=x \; \mbox{{ and}}\; l(r_1r_2;x)=l(r_1;l(r_2;x))$ with $x\in E$.

For spin vectors we have  $(E,l)=(\R^3,\lv)$
where 
$\lv(r;S):=rS$ and $r\in SO(3)$.
For a  spin tensor $M$ we invoke $(\Et,\lt)$ where
$\lt(r;M):=rMr\tran$ and $M\in E_t,r\in SO(3)$.
These are examples of the flexibility in the choice of $(E,l)$ mentioned in the Introduction.  

In this setting, over one turn, a field $f$  becomes the field $f'$ where \\
$ f'(\phi):=l(A(j\inv(\phi));f(j\inv(\phi))) $
or:
$ f \mapsto f'= l(A \circ j \inv; f \circ j \inv) $.
By definition, an invariant field maps into itself (the whole field) i.e., $f'= f $.
Then  invariance implies that $f(j(\phi) ) = l(A(\phi); f(\phi))$.
This, of course, reproduces the definitions of invariance of the ISF and the ITF
given earlier.
Appropriate  $SO(3)$-spaces can be introduced for handling the density matrices and
other objects related to spin.

With the language of $SO(3)$-spaces we are now in a position to
understand the relationship between invariant fields and
certain subgroups of $SO(3)$ as well as to explore the relationships
between different invariant fields and to classify them. Moreover we can associate invariant fields with 
certain invariant sets. We call our
approach the Technique of Association following its origins in Bundle
Theory and in 
%connection between our invariant fields and the cross sections on 
the so-called associated bundles lying behind our
structures.
\vspace{-0.0mm}
\section{\label{sec:NF}Normal Forms and the Normal Form Theorem (NFT)}
Spin motion can in general look complicated, especially close to spin-orbit resonances. \cite{BEH,arXiv2}
Nevertheless it can often be made to look simple by a proper continuous choice of a coordinate system
$T\in{\cal C}(\Td,SO(3))$
for the spin at each point on the torus.
Then we have 
$A'(\phi): =T^t( j(\phi))A(\phi)T(\phi)$ and within $T$ a spin vector $S$ is $S' = T^t(\phi)S$.
If the unit-length third axis of $T$ is the ISF, we call it an invariant frame field (IFF).  The motion of $S'$ is then a simple precession around the ISF
and $A'(\phi) \in SO(2)$. Away from orbital resonance, the other two axes of the IFF can usually be chosen so that 
the rate of precession of $S'$ in the IFF is independent of the orbital phases. This rate of precession is 
the ADST
% amplitude-dependent spin tune (ADST)  
and this leads to the definition of spin-orbit resonance. See Refs.~\cite{BEH} and \cite{arXiv2} for details.

In general we can classify the $A'$ according to their membership of subgroups 
{ $H$ of $SO(3)$ and we then call an $A'$ an $H$-normal form}. 
%This generalizes the concept of IFF from $SO(2)$ to $H$.}

This brings us to the Normal Form Theorem. We do not state it explicitly here but as explained in Refs.~\cite{IPAC14}--\cite{arXiv2} the 
NFT exploits the structures of $SO(3)$-spaces and it shows 
 { that an invariant field can be associated with a so-called {\em isotropy group} $H$ 
which gives us a particular $H$-normal form}.

Thus 
the ISF is associated with the subgroup $SO(2)$ of $SO(3)$.
Off orbital resonance, the ITF is  { generically}
associated with the subgroup $SO(2) \bowtie Z(2)$ 
of  $SO(3)$ where $\bowtie$ denotes a Zappa-Sz{\'e}p product.
The Normal Form Theorem shows how the notion of IFF  associated with $SO(2)$ can be 
generalized 
to apply to other isotropy groups, thereby providing a new view of the IFF, and it also
 provides a protocol for constructing further invariant $(E,l)$-fields. Moreover, it leads naturally to the 
Decomposition Theorem.
\vspace{-0.0mm}
\section{\label{sec:DT}The Decomposition Theorem (DT)}
Since invariant fields are tied to subgroups of $SO(3)$, these subgroups can be used to classify and relate 
invariant fields. For this we exploit the Decomposition Theorem. Again, we do not state the theorem,
but as explained in Refs.~\cite{IPAC14}--\cite{arXiv2}, it shows how, when $\Td \times E$ has been decomposed into certain 
disjoint invariant sets, the decompositions (the sets) in 
the same or different $(E,l)$ can be related and classified. In fact we have the powerful result that if the isotropy groups tied to two invariant fields are conjugate, then 
the two invariant fields are related by a homeomorphism. Moreover, if we know one invariant field, we can use homeomorphisms 
to construct others.

As an example we can use the relationship between the {
isotropy groups} tied to the ISF and ITF respectively to construct the ITF
from the ISF: $ f_t  = \pm   {\sqrt{\frac{3}{2}}} \left \{ \hat n {\hat n}\TR  - \frac{1}{3} I \right \} $.
In other words we can generate (say) the ITF from the ISF without recourse to physics! --- we have a machine to generate
non-arbitrary invariant fields. 
%But are they always physically relevant?  
Other examples of such generation can be found in Refs.~\cite{IPAC14}--\cite{arXiv2}.
\vspace{-0.0mm}
\section{\label{sec:IRT}The Invariant Reduction Theorem (IRT) and the Cross Section Theorem (CST)}
Our next offering is the Invariant Reduction Theorem.
\cite{IPAC14,arXiv1,arXiv2} 
{ In Section 8.7 of Ref.~\cite{arXiv2} we introduce a 
1-turn map $\hat{\cal P}$ on ${\Ed}:=\Td\times SO(3)$ and a special 
subset, $\Edh[f]$ of $\Ed$ depending on a field $f$. Then $f$ is an invariant 
field iff $\Edh[f]$ is invariant under ${\hat{\cal P}}$.

Thus
the important question of the existence of an invariant field, and of the ISF in particular, reduces to finding $f$ such that $\Edh[f]$ is 
invariant under $\hat{\cal P}$ and so
$\Edh[f]$ is the tool that we will use.}

The CST is even more technical than the previous three theorems and it
goes beyond the scope of this short summary. Suffice it to say that
the CST deals with so-called cross sections. { A cross section is the right-inverse}
{ of the natural
projection of $\Edh[f]$ into $\Td$ when $f$ is an
invariant field}.  For example, { if $f$ is an ISF then
there is a cross section iff there is an IFF whose third column is $f$. Thus
cross-sections give} new insights into ISF's and ITF's and
the CST provides an additional route for examining the question of
the existence of the ISF.  The reader should consult
Refs.~\cite{IPAC14}--\cite{arXiv2} for details.
\vspace{-0.0mm}
\section{\label{sec:bundle} Underlying Bundle Theory}
While it has not been necessary above to become immersed in the bundle-theoretic basis of the ToA
it is still appropriate to mention it since it supplies a steady flow of ideas. 
More on the mathematics can be found in
Refs.~\cite{Fe,HK} and \cite{Hu}  and in Refs.~\cite{IPAC14}--\cite{arXiv2} and \cite{HE1}.
We can also  draw analogies with the way that 
bundles are used in gauge theories for elementary particles
{ and see how
path lifting to parallel-transport motions can reproduce the T-BMT equation.}

The ``unreduced'' principal bundle underlying our formalism
is a product principal $SO(3)$-bundle with base space $\Td$.
For the ``unreduced'' principal bundle underlying a gauge theory the
base space is a patch of space-time and the fibres 
are elements of the gauge group.   
The transformation rule  $A'(\phi): =T^t( j(\phi))A(\phi)T(\phi)$ has its bundle counterpart
in a transformation rule under the
$SO(3)$ gauge-transformation group \cite{Hu}
of the unreduced principal bundle.

Every $(E,l)$ in the formalism uniquely determines an ``associated
bundle'', relative to the unreduced bundle. In our formalism the
invariant $(E,l)$-fields are the nontrivial data of invariant cross
sections of associated bundles. This is analogous to the case of gauge
theories where the matter fields carry the data of cross sections of
associated bundles. With this we have ``geometrized'' the invariant
fields and provided a new view of invariant fields.
\section{\label{sec:summary} Summary and Plans}
We have a new, concise and powerful formalism, based on the concept of
$SO(3)$-spaces $(E,l)$, for defining, generalizing, creating and
classifying invariant fields in storage rings using tools inspired by
Bundle Theory. 
%(-- gauge theories also exploit bundles).
In fact by their origin in bundle theory, invariant fields and IFF's are rather deep concepts.\cite{HE1}
%We will persue the analogy with bundles to find parallels useful for studying  invariant fields. 
Furthermore we have a new tool for studying the question of the existence of invariant $(E,l)$-fields and in particular the existence of the ISF.
With no ISF, there can be no equilibrium polarization.
We believe that this is the first application of these methods \cite{Zi,Fe,HK} in accelerator physics.

\section*{Acknowledgments}
 Work has been supported by DOE under DE-FG-99ER41104
 and by DESY. We thank Dan T. Abell for help with preparing figures for the talk itself.


\begin{thebibliography}{0} % Use for 10-99 references
\bibitem{BEH}
D.P. Barber, J.A. Ellison and K. Heinemann,
%\textit{Quasiperiodic spin-orbit motion and spin tunes in storage rings},
\textit{Phys. Rev. ST Accel.
Beams} {\bf 7}, 124002 (2004).
%
%
\bibitem{HH}
K. Heinemann and G.H. Hoffstaetter, \textit {Phys. Rev.} {\bf E 54} (4), 4240 (1996).
%{\em Tracking algorithm for the stable spin polarization field in storage rings using
%stroboscopic averaging}, Phys.Rev. {\bf E 54} (4), 4240 (1996).
%
%
\bibitem{IPAC14}
K. Heinemann, D.P. Barber, J.A. Ellison and M. Vogt,
 \textit{Proc. 2014 Int. Particle  Accelerator Conference}, Dresden,
         Germany, June 2014, JACOW, (2014).
%
%
\bibitem{arXiv1}
K. Heinemann, D.P. Barber, J.A. Ellison and M. Vogt,
arXiv:1409.4373v2  (physics.acc-ph, math-ph, math.MP) (2014).
Submitted for publication.
%
%
\bibitem{arXiv2}
K. Heinemann, D.P. Barber, J.A. Ellison and M. Vogt
arXiv:1501.02747v1 (physics.acc-ph, math-ph, math.MP) (2015).
Submitted for publication.
%
%
\bibitem{Zi} 
R.J. Zimmer,
Ergodic theory and the automorphism group of a G-structure,
\textit{Proc. of a conference in honor of George W. Mackey}, Berkeley, USA, 1984.
Mathematical Sciences Research Institute Publications, (Springer-Verlag,
1987), p. 247.
%
%
\bibitem{Fe}
R. Feres, \textit{Dynamical systems and semisimple groups: an introduction},
(Cambridge University Press, Cambridge, 1998).
%
\bibitem{HK}
R. Feres and  A. Katok,
\textit{Handbook of dynamical systems Vol. 1A}.
eds. B. Hasselblatt and A. Katok,
(North-Holland, Amsterdam, 2002).
%
%
\bibitem{HE1}
K. Heinemann, \textit{Two Topics in Particle
Accelerator Beams}, Doctoral Thesis, University of New Mexico (2010).
%
%
\bibitem{BV2}
D.P. Barber and  M. Vogt,
The Invariant Polarisation-Tensor Field for Spin-1 Particles
        in Storage Rings, 
\textit{Proc. 18th Int. Spin Physics Symposium},
Charlottesville, USA, October 2008, (AIP proceedings 1149, 2008).
%
%
%\bibitem{CFS}
%I.P. Cornfeld, S.V. Fomin and Y.G. Sinai, ``Ergodic Theory'', Springer (1982).
%
%
\bibitem{Hu}
D.  Husemoller, \textit{Fibre Bundles}, 3rd edn. (Springer-Verlag, New York,
1994).
%
%
\end{thebibliography}
\end{document}